\documentstyle[aps,twocolumn]{revtex}

\input psfig

\begin{document}

\title{Comment on ``New Vortex-Matter Size Effect Observed in Bi$_2$Sr$_2$CaCu$_2$O$_{8+\delta}$"}  
\author{Y. Kopelevich$^+$ and P. Esquinazi$^*$}\address{$^+$Instituto
de Fisica,  Unicamp, 13083-970 Campinas, Sao Paulo, Brasil}  
\address{$^*$Abteilung Supraleitung und  Magnetismus, Universit\"at
Leipzig, Linn{\'e}str. 5, D-04103 Leipzig, Germany} 

\maketitle 
\vspace{-13mm}

The Letter of Wang et al. \cite{wan} is related to the recently discovered 
vanishing of the second magnetization peak (SMP) in high-$T_c$ superconductors 
reducing the sample size \cite{kope1,kope2,gal}. This phenomenon has been 
interpreted \cite{kope1,kope2,gal} in terms of thermomagnetic instability (TMI) effects. 
The TMI approach has also been adopted by Wang et al. in their previous work 
\cite{wan2} on the critical sample size temperature dependence $R_{cr}(T)$ measured 
in  Bi$_2$Sr$_2$CaCu$_2$O$_{8+\delta}$ (Bi2212) single crystals. In contrast, in the 
recently published Letter \cite{wan} the authors claimed that the $R_{cr}(T)$ obtained 
for Bi2212 crystals is at odds with the TMI model and proposed an alternative 
explanation of the results assuming that the SMP vanishes as the sample width 
(or diameter) becomes smaller than the in-plane vortex correlation length 
$R_c^{2D}(T) \sim J_c(T)^{-1/2}$, where $J_c(T)$ is the critical current density.
We demonstrate below that the authors conclusion is misleading.

First of all, we stress that the authors used criterion for the occurrence of the SMP 
$R(T) \ge  R_c^{2D}(T) \equiv R_{cr}(T)$ contradicts the low-temperature results. 
Indeed, the SMP in Bi2212 does not occur at temperatures below $\sim 15~$K$\ldots$20~K 
in crystals of  mm size, i.e. with $R \gg R_{cr}(T)$ (see~\cite{kope2} and refs. therein).

Next, the reported $R_{cr}(T)$ \cite{wan} agrees very well with the TMI scenario. 
Figure 1(a) shows the $R_{cr}(T)$ data from \cite{wan}. The solid line in 
Fig.1(a) is obtained from the equation for the critical sample size according 
to the TMI theory \cite{swa}
\begin{equation}
R_{cr} \simeq (-100 \pi C(T) / 16 J_c(T) \partial J_c/\partial T)^{1/2}\,,
\end{equation}
where we assumed for the specific heat $C \propto T^3$ and  $J_c(T) \propto \exp(-T/T_0)$ 
\protect\cite{wan}.
As Fig. 1(a) demonstrates Eq. (1) fits
nicely the experimental data. For a quantitative comparison precise knowledge of 
$C(T)$ and $J_c(T)$ for the measured samples is required.

Moreover, $R_{cr}(T)$ obtained for Bi2212 \cite{wan}
 is similar to that reported for Nb films \cite{esq}, where it has unambiguously been 
demonstrated that $R_{cr}(T)$ separates the adiabatic and isothermal critical states. 
This similarity is illustrated in Fig.1(b) where the effective critical sample size 
$s_{cr} = (wd/2)^{1/2}$ \cite{kope1,kope2,gal}  is plotted vs. $T/T_c$ for 
both superconductors, with $w$ and $d$ being the sample width (or diameter) and thickness. 

In \cite{kope2} we argued that TMI in Bi2212 is triggered by plastic vortex motion 
which ceases for $T < 15~$K$\ldots$20~K, providing us a natural explanation for the 
SMP vanishing at low temperatures. Summarizing, the results reported in \cite{wan} 
are in good agreement with the TMI scenario of the sample-size-dependent SMP in Bi2212
and partially contradict the authors conclusion.

\begin{figure}
\centerline{\psfig{file=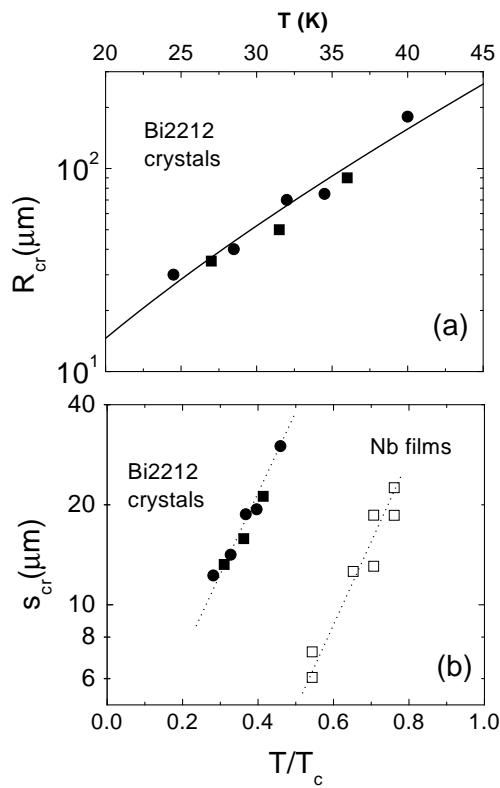,height=4.2in}}
\caption{(a) $R_{cr}(T)$ from \protect\cite{wan}. The solid line is obtained from Eq. (1)
$R_{cr}(T) = A T^{3/2} \exp(T/T_0)$ with $A = 0.043~\mu$m/K$^{3/2}$ and $T_0 = 15~$K. 
(b) Effective sample size vs. reduced temperature
obtained for Bi2212 crystals \protect\cite{wan} ($T_c = 87~$K) 
and Nb-films \protect\cite{esq} ($T_c = 9.2~$K). Dashed lines are only a guide.}
\end{figure}

\end{document}